\documentclass[preprint,prc,aps,epsf]{revtex4}
\usepackage{graphicx}
\usepackage{epstopdf}
\usepackage{epsfig}

\begin{document}

\title{Nuclear effects in high-energy neutrino interactions}

\author{Spencer R. Klein, Sally A. Robertson}
\affiliation{Lawrence Berkeley National Laboratory, Berkeley CA, 94720, USA}

\author{Ramona Vogt}
\affiliation{Nuclear and Chemical Sciences Division,
  Lawrence Livermore National Laboratory, Livermore, CA 94551, USA}
\affiliation{Physics Department, University of California at Davis, Davis, CA,
  95616, USA}

\date{\today}

\begin{abstract}
%SRK added Baikal below and capitalized 'T' in KM3NeT
Neutrino telescopes like IceCube, KM3NeT and Baikal-GVD  offer physicists the opportunity to study neutrinos with energies far beyond the reach of terrestrial accelerators.  These neutrinos are used to study high-energy neutrino interactions and to probe the Earth through absorption tomography.   Current studies of TeV neutrinos use cross sections which are calculated for free nucleons with targets which are assumed to contain equal numbers of protons and neutrons.  

Here we consider modifications of high-energy neutrino interactions due to two nuclear effects: modifications of the parton densities in the nucleus, referred to here as shadowing, and the effect of non-isoscalar targets, with unequal numbers of neutrons and protons.  Both these effects depend on the interaction medium.  Because shadowing is larger for heavier nuclei, such as iron, found in the Earth's core, it introduces a zenith-angle dependent change in the absorption cross section.  These modifications increase the cross sections by 1-2\% at energies below 100 TeV (antishadowing), and reduce it by 3-4\% at higher energies (shadowing).  

Nuclear effects also alter the inelasticity distribution of neutrino interactions in water/ice by increasing the number of low inelasticity interactions, with a larger effect for $\nu$ than $\overline\nu$.   These effects are particularly large in the energy range below a few TeV.  These effects could alter the cross sections inferred from events with tracks originating within the active detector volume as well as the ratio $\nu/\overline\nu$ inferred from inelasticity measurements.

The uncertainties in these nuclear effects are larger than the uncertainties on the free-proton cross sections and will thus limit the systematic precision of future high-precision measurements at neutrino telescopes. 

\end{abstract}

\maketitle

\section{Introduction}
% SRK removed 'like IceCube' to keep more generic, and so the first and second paragraphs start more differently. 
Neutrino telescopes have observed neutrinos with energies well above $10^6$~GeV \cite{Aartsen:2016xlq}.  Future  experiments will use radio-Cherenkov techniques to search for neutrinos with energies up to $10^{20}$ eV and beyond \cite{Ackermann:2019cxh} due to interactions of ultra-high energy cosmic-rays with the cosmic microwave background radiation.  These neutrinos are important probes of the cosmos; Astrophysical neutrinos should point back to the locations of high-energy cosmic-ray accelerators in the universe \cite{Aartsen:2019mbc} while measurements of the diffuse, $4\pi$, flux are sensitive to the properties of the accelerators \cite{Aartsen:2015ivb}.  Atmospheric neutrinos are sensitive to the composition of cosmic rays and also to some aspects of hadron physics \cite{Aartsen:2015xup,Aartsen:2018vez}.  High-energy neutrinos are also used to study neutrino interactions and to search for beyond the standard model (BSM) physics \cite{Aartsen:2017ibm}.

%SRK Remove "Neutrino Telescopes like" to further differentiate openings of first and second paragraphs. %RV-- changed 'have' to 'has' for subject-verb agreement
The 1 km$^3$ IceCube neutrino telescope \cite{Halzen:2010yj} has collected large samples of neutrino events, sometimes comprising more than 500,000 neutrino events \cite{Aartsen:2017mau}.   Optical sensors observe the Cherenkov radiation from relativistic charged particles produced in the interactions.  These events may be through-going muons from neutrino interactions outside the detector, or starting events, neutrino interactions within the detector.  More complex topologies are also possible. Through-going muon analyses usually bin the data by muon energy (a proxy for neutrino energy) and zenith angle then fitting these data to models making different assumptions about the astrophysical neutrino flux and angular distribution (isotropicity), possibly including neutrino propagation and interaction, including BSM interactions.   The muon energy in the detector is inferred by measurement of its specific energy loss, $dE/dx$ \cite{Abbasi:2012wht}.   

At energies above a few TeV, absorption in the Earth alters the zenith angle distribution.  The absorption correction should be modeled with a precision similar to the other uncertainties in the zenith angle distribution, reaching a few percent.  It is also important to accurately model the neutrino inelasticity distribution since it is important for inferring the neutrino energy spectrum from the muon energy spectrum.   In charged-current $\nu_\mu$ and $\overline\nu_\mu$ starting event interactions, both the hadronic cascade and the outgoing lepton are observed. Thus inelasticity becomes a third variable in astrophysical neutrino flux fits.  The energy of the hadronic cascades is inferred through calorimetry \cite{Aartsen:2013vja}.  

Natural neutrinos are also used to study neutrino interactions at energies above those that are accessible at accelerators, where the highest available neutrino energy is 500 GeV, although this range will be extended by the planned FASER$\nu$ experiment located at the Large Hadron Collider \cite{2019arXiv190802310F}.  IceCube has measured the neutrino-nucleon cross sections for $\nu_\mu$ \cite{Aartsen:2017kpd} and for a primarily $\nu_e$ mixture \cite{Yuan:2019wil,Bustamante:2017xuy}.  They have also measured the neutrino inelasticity distribution \cite{Aartsen:2018vez} and provided interaction data used  to produce a tomographic image of the mass distribution within the Earth \cite{Kotoyo,Donini:2018tsg}.  These data have been used to search for BSM processes \cite{Ellis:2016dgb,Klein:2019nbu}.  The cross section and Earth tomography analyses rely on measurements of neutrino absorption within the Earth while the inelasticity measurements and most BSM studies use direct observations of neutrino interactions  in Antarctic ice.   The tomography effort is noteworthy because seismographic measurements of the Earth's density are already quite precise \cite{densityuncertainty}.  Therefore any neutrino-based measurement must have a comparably small statistical and systematic errors to be competitive.   For example, in Ref.~\cite{Aartsen:2017kpd} IceCube assigned a 1\% systematic uncertainty to account for uncertainties in the Earth's mass distribution.  

These studies assume that high-energy neutrinos interact with isospin-invariant, unshadowed targets via Deep inelastic Scattering (DIS).   The purpose of this paper is to examine how nuclear corrections (compared to an isospin-invariant non-shadowed reference) to DIS cross sections and inelasticity distributions affect the assumptions used in neutrino-telescope analyses.  We focus on energies above 1~TeV. 

%SRK added '(NLO)' here, the first time it is used
The reference cross sections for these studies are next-to-leading order (NLO) perturbative QCD calculations \cite{CooperSarkar:2011pa,Connolly:2011vc}.   Because neutrino interactions involve $W^\pm$ and $Z^0$ exchange, higher-order corrections to the cross sections should be small and the dominant uncertainties should arise from the uncertainties in the parton distribution functions (PDFs).  %The largest PDF uncertainty is at low Bjorken-$x$, corresponding to high neutrino energies.
%SRK changed 'shown' to 'found'
The uncertainties on these cross sections due to the proton PDFs was found to be 5\% for $E_\nu$ from 50~GeV to $10^{9}$~GeV, rising significantly at higher energies \cite{CooperSarkar:2011pa}.

These calculations have some limitations.  While they assume free-nucleon targets, the Earth contains significant contributions from elements as light as hydrogen to as heavy as iron.  Shadowing may be significant for the heavier nuclei.  The calculations assume that the targets are isoscalar, with equal numbers of protons and neutrons.   However, H$_2$O is 62.5\% protons.  Some other elemental components of the Earth are neutron rich.  The
GENIE model \cite{Andreopoulos:2015wxa}, which is commonly used to model neutrino interactions at energies up to a few hundred GeV, includes non-isoscalar targets and some other nuclear effects \cite{Andreopoulos:2015wxa}.  At these lower energies, neutrino interaction phenomenology is quite rich \cite{Andreopoulos:2019gvw}.

This paper will consider how shadowing and non-isoscalar targets affect neutrino absorption in the Earth and the observed inelasticity distribution of interactions in ice.  We explore the change to neutrino absorption for different paths through the Earth, accounting for the different elemental components of the core, mantle and crust; determine the nuclear modifications for the elements in different chords through the Earth ({\it i.e.}\ different zenith angles); and calculate the overall change in cross sections.   We use a leading-order pQCD calculation, forming a ratio to a parallel calculation assuming an unshadowed, isoscalar, deuterium target.  We do not consider initial-state phenomena such as the
colored glass condensate (CGC) \cite{Gelis:2010nm}, or other factors that might
lead to larger changes in the cross sections.  
%SRK made some changes in the sentence below %RV made more
The presence of a CGC was predicted to greatly reduce the neutrino-nucleon interaction cross sections at energies above $\sim 10^{10}$ GeV \cite{Henley:2005ms} because of the large change in the gluon density.  
%SRK this clause doesn't fit the flow here and seems unneeded.  People who know what a CGC is know about dipole calculations: required to calculate the dipole cross section.  
The calculation in Ref.~\cite{Henley:2005ms} considered free nucleons.  However, heavier nuclear targets increase the minimum Bjorken-$x$ required for the onset of the CGC regime as $A^{1/3}$ \cite{Kovchegov:2012csa} where $A$ is the atomic number. In the case of iron, $A^{1/3}\approx 3.8$.  Thus, in addition to reducing the overall neutrino cross section, a CGC initial state would also alter in the zenith-angle dependence of neutrino absorption in the Earth, in particular, a larger reduction in cross section in the Earth's core. 
%SRK added references & discussion of other two calculation %RV made a few edits, would prefer to use GeV or TeV units, better point of reference to NP than eV 
Another calculation considered the effects of nonlinear QCD evolution, at sufficiently high parton densities for recombination to be important, and found a smaller, factor-of-2, reduction in the cross section \cite{Goncalves:2013kva} at energies as low as $10^{4}$~TeV.  This result seems to exhibit at least mild tension with the IceCube $\nu_\mu$ analysis \cite{Aartsen:2017kpd}.  A third calculation found a similar reduction at higher energies, above $10^{8}$~TeV \cite{Albacete:2015zra}.  They noted that, in this regime, nonlinearities lead to significant uncertainties in the cross section. 

We consider the effects on the cross section and inelasticity distribution for neutrinos interacting in ice or water, particularly the effects on hydrogen with a nucleus of a single proton.   We  discuss how these nuclear effects, and their uncertainties, affect measurements that can be performed using ice-based neutrino telescopes. 

\section{Calculations of nuclear effects}

%SRK removed 'next-to-leading-order' since NLO is now defined above. 
Both charged lepton and neutrino interactions in nuclei in deep-inelastic scattering (DIS) measurements show modifications of the quark parton densities in the medium, even for nuclei with masses as low as $A = 4$.  Here we give the neutrino-nucleus cross sections for charged and neutral currents for non-isoscalar targets including nuclear modifications of the parton densities (nPDFs), relative to a deuteron.  Although our calculation is leading order (LO), we use NLO nPDFs and PDFs to obtain a first estimate of the effect on absorption of neutrinos by the Earth.  There are few strictly LO proton PDFs and those that exist do not include error sets.  In addition, the latest nPDF set, EPPS16 \cite{EPPS16} only has a NLO set available.  We have checked the proton to deuterium ratio at LO calculated with both CTEQ61L (a fully LO calculation) and with CTEQ6M \cite{cteq6m} (LO cross sections and NLO proton PDFs) and found agreement between the two calculations on the sub-percent level.  We chose these two sets because they were the CTEQ sets used in the global analyses of the EPS09 nPDF sets at LO and NLO respectively.  (We note that the nature of the gluon densities in these sets may lead to some differences for gluon-dominated processes.  This would not be the case for other nPDF sets based on proton parton densities with similar small $x$ gluon behavior \cite{RVppb}.)  This justifies using a LO calculation to test the modifications in question using calculated ratios.  (We remark that Ref.~\cite{Dev} calculated neutrino-nucleon scattering at LO with leading, next-to-leading and next-to-next-to-leading order proton PDFs and found only small differences.  They also noted that their LO cross sections were within a few percent of a previous NLO calculation \cite{Dev}.)

Neglecting the longitudinal structure function, $F_L$, the charged current cross section for neutrino- or antineutrino-proton interactions is
\begin{eqnarray}
\frac{d^2\sigma^{{\rm CC}}(\nu(\overline \nu)p)}{dx dQ^2} = \left[\frac{G_F^2 M_W^4}
{4\pi x (Q^2 + M_W^2)^2} \right] \sigma^{{\rm CC}}(\nu(\overline \nu)p)
\label{eq:sig_cc}
\end{eqnarray}
where
\begin{eqnarray}
\sigma^{{\rm CC}}(\nu p) & = & Y_+ F_{2p}^{\nu {\rm CC}}(x,Q^2) +
Y_- xF_{3p}^{\nu {\rm CC}}(x,Q^2)
\,\, , \label{eq:sig_cc_nup} \\
\sigma^{{\rm CC}}(\overline \nu p) & = & Y_+ F_{2p}^{\overline \nu {\rm CC}}(x,Q^2) -
Y_- xF_{3p}^{\overline \nu {\rm CC}}(x,Q^2)
\,\, , \label{eq:sig_cc_nubp} \\
\end{eqnarray}
with $Y_\pm = 1 \pm (1 - y)^2$.  Here $G_F$ is the Fermi constant, $M_W$ is the $W^\pm$ boson mass, $Q^2$ is the square of the momentum transferred
from the neutrino to the proton.  The fraction of the proton momentum carried by a quark is $x = Q^2/(2m (E_\nu- E_\nu'))$ with proton mass $m$, incoming and outgoing neutrino energies $E_\nu$ and $E'_\nu$ respectively, and inelasticity $y = (E_\nu-E'_\nu)/E_\nu$.

The structure functions for the proton, $F_{2p}^{\nu {\rm CC}}$ and $xF_{3p}^{\nu {\rm CC}}$, for an exchanged $W^+$ are
\begin{eqnarray}
F_{2p}^{\nu {\rm CC}} & = & 2x (d + \overline u + s + \overline c + b) \, \,
\label{F2p_nuCC} \, \, , \\
x F_{3p}^{\nu {\rm CC}} & = & 2x (d - \overline u + s - \overline c + b) \, \,
\label{F3p_nuCC} \, \, . 
\end{eqnarray}
Here we suppress the dependence of the structure functions and the individual quark parton densities on $x$ and $Q^2$ to be more concise.  Note that $u$, $d$, $s$, $c$ and $b$ refer to the up, down, strange, charm and bottom quark distributions.  The up and down distributions include contributions from the valence quarks that define the proton identity, $u_V$ and $d_V$, while the perturbatively-generated sea quark and antiquark distributions can be referred to as $q_s$ or $\overline q$ interchangeably.  We assume $q_s = \overline q_s = \overline q$ since these distributions are produced pairwise through gluon splitting and do not contribute to the baryon number or charge of the hadron, unlike the valence quark distributions.  These sea distributions dominate the quark PDFs at low momentum.  We also define $d = d_V + d_s = d_V + \overline d$, $u = u_V + u_s = u_V + \overline u$, $s = \overline s$, $c = \overline c$ and $b = \overline b$. Likewise, for an exchanged $W^-$, the structure functions for charged current interactions of antineutrinos with a proton are
\begin{eqnarray}
F_{2p}^{\overline \nu {\rm CC}} & = & 2x (u + \overline d + c + \overline s +
\overline b) \, \,
\label{F2p_nubCC} \, \, , \\
x F_{3p}^{\overline \nu {\rm CC}} & = & 2x (u - \overline d - \overline s + c -
\overline b) \, \,
\label{F3p_nubCC} \, \, . 
\end{eqnarray}
To go from a proton target to a nuclear target, we have to define the structure functions for charged current neutrino interactions on a neutron.  We take $u^p = d^n$, $d^p = u^n$, $\overline u^p = \overline d^n$ and $\overline d^p = \overline u^p$.  We make the distinction for the light quark sea because the $\overline u$ and $\overline d$ distributions were found to be
different in studies of Drell-Yan dilepton production in $p+p$ and $p+d$ interactions \cite{Baldit:1994jk,Towell:2001nh,Nagai:2018scg}.  This difference has been incorporated into global analyses of the proton PDFs since the MRS(A) \cite{Martin:1996as} and CTEQ4 \cite{Lai:1996mg} sets.  Writing the neutron structure functions in terms of the proton parton densities, we have
\begin{eqnarray}
F_{2n}^{\nu {\rm CC}} & = & 2x (u + \overline d + s + \overline c + b) \, \,
\label{F2n_nuCC} \, \, , \\
x F_{3n}^{\nu {\rm CC}} & = & 2x (u - \overline d + s - \overline c + b) \, \,
\label{F3n_nuCC} \, \, . \\
F_{2n}^{\overline \nu {\rm CC}} & = & 2x (d + \overline u + c + \overline s +
\overline b) \, \,
\label{F2n_nubCC} \, \, , \\
x F_{3n}^{\overline \nu {\rm CC}} & = & 2x (d - \overline u - \overline s + c -
\overline b) \, \,
\label{F3n_nubCC} \, \, . 
\end{eqnarray}
%SRK changed 'generic neutrino' to 'nucleus' %RV added back neutrino because charge lepton structure functions aren't identical
We write the structure functions for neutrino-nucleus interactions with contributions from $Z$ protons and $N$ neutrons in a nucleus of mass number $A = Z + N$, leaving out, for the moment, the modifications due to the nuclear medium, as
\begin{eqnarray}
F_{2A}^{\nu {\rm CC}} & = & \frac{1}{A} (Z F_{2p}^{\nu {\rm CC}} + N F_{2n}^{\nu {\rm CC}}) \,\, ,
\\
x F_{3A}^{\nu {\rm CC}} & = & \frac{1}{A} (Z x F_{3p}^{\nu {\rm CC}} +
N x F_{3n}^{\nu {\rm CC}}) \,\, .
\end{eqnarray}
For an isoscalar target, $Z = N = A/2$, we have {\it e.g.} $F_{2A}^{\nu {\rm CC}} = 0.5 (F_{2p}^{\nu {\rm CC}} + F_{2n}^{\nu {\rm CC}}) = x (u + \overline u + d + \overline d + 2 \overline s + 2c + 2 \overline b)$, the usual definition for these calculations. The results for the charged current structure functions with a nuclear target are
\begin{eqnarray}
F_{2A}^{\nu {\rm CC}} & = & \frac{2x}{A} (Z(d + \overline u) + N(u + \overline d) +
A(s + \overline c + b)) \, \,
\label{F2A_nuCC} \, \, , \\
x F_{3A}^{\nu {\rm CC}} & = & \frac{2x}{A} (Z(d - \overline u) + N(u - \overline d)
+ A(s - \overline c + b)) \, \,
\label{F3A_nuCC} \, \, . \\
F_{2A}^{\overline \nu {\rm CC}} & = & \frac{2x}{A} (Z(u + \overline d)
+ N(d + \overline u) + A(c + \overline s +
\overline b)) \, \,
\label{F2A_nubCC} \, \, , \\
x F_{3A}^{\overline \nu {\rm CC}} & = & \frac{2x}{A} (Z(d - \overline u)
+ N(d - \overline u) + A(c - \overline s -
\overline b)) \, \,
\label{F3A_nubCC} \, \, . 
\end{eqnarray}

To include nuclear modifications of the parton densities in the nucleus, we introduce the shadowing ratios, $R_i(x,Q^2,A)$, discussed below.  We assume that they are distinct for each quark flavor, as well as differentiating between valence and sea contributions for up and down quarks.  The nPDFs are determined from global analyses of experiments with nuclear targets, including Drell-Yan and nuclear DIS with charged leptons and neutrinos (in some cases).  The various contributing processes allow separation of effects on the valence and sea distributions.  These analyses will be discussed further shortly.

We use the EPS09 \cite{EPS09} and EPPS16 \cite{EPPS16} sets at next-to-leading order for the modifications.  Thus, in Eq.~(\ref{eq:sig_cc}), we replace $\sigma^{{\rm CC}}(\nu(\overline \nu)p)$ by $\sigma^{{\rm CC}}(\nu(\overline \nu)A)$ and the proton structure functions in Eqs.~(\ref{eq:sig_cc_nup}) and (\ref{eq:sig_cc_nubp}) with the non-isoscalar structure functions including these nuclear modification ratios:
%SRK added some \! so formulae don't overlap with equation numbers
\begin{eqnarray}
F_{2A}^{\nu {\rm CC}} &\! =\! & \frac{2x}{A} (Z(R_{d_V} d_V\! +\! R_{\overline d} \overline d
\!+\! R_{\overline u}\overline u)\! +\! N(R_{u_V}u_V +R_{\overline u} \overline u\! +\!
R_{\overline d}\overline d) \!+\!
A(R_s s\! +\! R_c \overline c\! +\! R_b b)) \, \,
\label{F2As_nuCC} \, \, , \\
x F_{3A}^{\nu {\rm CC}} &\! =\! & \frac{2x}{A} (Z(R_{d_V}d_V\! +\!\! R_{\overline d}\overline d
\!-\! R_{\overline u}\overline u)\! +\! N(R_{u_V}u_V + R_{\overline u}\overline u
\!-\! R_{\overline d}\overline d)
\!+\! A(R_s s \!-\! R_c \overline c\! +\! R_b b)) \, \,
\label{F3As_nuCC} \, \, . \\
F_{2A}^{\overline \nu {\rm CC}} &\! =\! & \frac{2x}{A} (Z(R_{u_V}u_V\! +\!
R_{\overline u}\overline u\! +\! R_{\overline d}\overline d)
\!+\! N(R_{d_V}d_V\! +\! R_{\overline d}\overline d \!+\! R_{\overline u}\overline u)
\!+\! A(R_c c\! +\! R_S \overline s \!+\! R_b \overline b)) \, \,
\label{F2As_nubCC} \, \, , \\
x F_{3A}^{\overline \nu {\rm CC}} &\! = \!& \frac{2x}{A} (Z(R_{d_V}d_V
\!+\! R_{\overline d}\overline d \!-\! R_{\overline u}\overline u)
\!+\! N(R_{d_V}d_V\! +\! R_{\overline d}\overline d\! -\! R_{\overline u}\overline u)
\! +\! A(R_c c\! -\! R_s \overline s \!-\! R_b \overline b)) \, \,
\label{F3As_nubCC} \, \, . 
\end{eqnarray}
We assume above that the sea quark modifications are identical for quarks and antiquarks, {\it i.e.} $R_{u_s} = R_{\overline u}$,
$R_{d_s} = R_{\overline d}$, $R_s = R_{\overline s}$ {\it etc.}

We now turn to neutral currents.  The cross sections are defined similarly
\begin{eqnarray}
\frac{d^2\sigma^{{\rm NC}}(\nu(\overline \nu)A)}{dx dQ^2} = \left[\frac{G_F^2 M_Z^4}
{4\pi x (Q^2 + M_Z^2)^2} \right] \sigma^{{\rm NC}}(\nu(\overline \nu)A)
\label{eq:sig_nc}
\end{eqnarray}
where
\begin{eqnarray}
\sigma^{{\rm NC}}(\nu A) & = & Y_+ F_{2A}^{\nu {\rm NC}}(x,Q^2) +
Y_- xF_{3A}^{\nu {\rm NC}}(x,Q^2)
\,\, , \label{eq:sig_nc_nuA} \\
\sigma^{{\rm NC}}(\overline \nu A) & = & Y_+ F_{2A}^{\overline \nu {\rm NC}}(x,Q^2) -
Y_- xF_{3A}^{\overline \nu {\rm NC}}(x,Q^2).
\,\, , \label{eq:sig_cc_nubA} 
\end{eqnarray}
We have now directly written the cross sections and structure functions in terms of nuclear mass number $A$.  In this case, there are additional $u$-like and $d$-like couplings on the structure functions \cite{coupling_refs}.  With $F_2$, the $u$-like couplings $a_u^2 + v_u^2$ multiply the up and charm parton densities while the $d$-like couplings $a_d^2 + v_d^2$ multiply the
down, strange and bottom parton densities.  In the case of $xF_3$, the $u$-like couplings are $2a_u v_u$ while the $d$-like couplings are $2a_d v_d$.  Recall that $a_u = 1/2$, $a_d = -1/2$, $v_u = 1/2 - (4/3)\sin^2\theta_W$ and $v_d = -1/2 + (2/3)\sin^2 \theta_W$.   The subtraction of the identical $q$ and $\overline q$ at the vertices for the $F_3$ structure function with isoscalar target and no shadowing leads to very simple structure in this case, depending only on the valence quarks with $F_{3A}^{\nu {\rm NC}} = F_{3A}^{\overline\nu {\rm NC}}$.  While we find that, in general, $F_{2A}^{\nu {\rm NC}} = F_{2A}^{\overline\nu {\rm NC}}$, there is no cancellation when $N \neq Z$ and the expression is therefore more complex.  Including the nuclear modifications, the structure functions for the neutral current interactions are
\begin{eqnarray}%missing bracket
\!F_{2A}^{\nu (\overline\nu) {\rm NC}}\! &\!= \!& \frac{1}{A} [(a_u^2 + v_u^2)[Z (R_{u_V}u_V  
+ 2R_{\overline u}\overline u + 2R_c c)\!  \\ \nonumber & &  \mbox{}  \qquad 
+ \! N\! (R_{d_V}d_V
+ 2R_{\overline d} \overline d + 2R_s s + 2 R_b b)] \\ \nonumber
&  & \mbox{} + (a_d^2 + v_d^2)[N
(R_{u_V}u_V + 2R_{\overline u}\overline u + 2R_c c)\!  %\\ \nonumber & &  \mbox{} \qquad 
+ \ \!Z\! (R_{d_V}d_V
+ 2R_{\overline d} \overline d + 2R_s s + 2 R_b b)]] \, \, \label{F2A_nc} \\
\!xF_{3A}^{\nu (\overline\nu) {\rm NC}}\! &\!=\! & \frac{2x}{A}[ a_u v_u (Z R_{u_V}u_V
+ NR_{d_V}d_V) +  a_d v_d (N R_{u_V}u_V
+ ZR_{d_V}d_V)] \, \, . \label{F3A_nc}
\end{eqnarray}

We now turn to nuclear shadowing parameterizations. We use the CT10 proton parton densities \cite{CT10} for the free nucleons in our further calculations.  We use the central set only and do not include uncertainties in the proton PDFs here to focus on the effect of the nuclear modifications.  We will, however, discuss the sensitivity of the inelasticity distributions to the proton PDF choice in Sec.~\ref{Sec:inelasticity}.

Both of the parameterizations of the nuclear modifications used in this work assume collinear factorization with DGLAP evolution.  Both sets are optimized assuming that there is no shadowing effect present in deuterium ($A = 2$, $Z=1$).  While shadowing may depend on where the probe impacts the nucleus, for example, closer to the `edge' of the nucleus where there might be only one nucleon in its path or more in the center where it may encounter multiple nucleons \cite{Kitagaki:1988wc,Emelyanov:1999pkc}, this is not taken into account.  Thus the parameterizations themselves are blind to the nuclear shape and density so a more loosely bound nucleus like $^6$Li, which might be described as an alpha particle ($^4$He) with two neutrons is treated the same way as a tightly bound nucleus such as $^{56}$Fe with two closed shells.  Or, nuclei with neutron skins might experience enhanced shadowing for protons with correspondingly weaker shadowing for neutrons. These effects are small for the nuclei commonly found in the Earth, and are not likely to affect our results. 

EPS09 \cite{EPS09} defines three different nuclear corrections at the initial scale $Q_0^2 = 1.69$~GeV$^2$: $R_V^A$ for both up and down valence quarks; $R_S^A$ for all sea quarks; and $R_G^A$ for gluons.
Fifteen fit parameters were employed, resulting in 30 error sets determined by varying each parameter by one standard deviation in each direction from its optimized value, in addition to the best fit, central set.  The nuclear dependence of each of the parameters is assumed to follow $(A/A_{\rm ref})^{p_i}$ where $A_{\rm ref} = 12$ and $p_i$ is a fit parameter.  Uncertainties on the individual quark ratios are calculated by summing the excursions of each of the error sets from the central value in quadrature.  The sets cover the range $1.3 < Q < 1000$~GeV and $10^{-6} < x < 1$.  Outside of these ranges, the value of the required ratio at the minimum $x$ or maximum $Q$ is returned.  

This set, developed before the LHC turned on, relied primarily on fixed-target DIS of electrons and muons from nuclear targets of He, Li, Be, C, Al, Ca, Fe, and Cu measured relative to scattering off deuterium \cite{20,21,22,23,26}. Drell-Yan studies from the Fermilab E772 \cite{E772} and E866 \cite{E866} experiments produced ratios of C/D, Ca/D, Fe/D (E772) and Fe/Be (E866) that could be used to separate valence from sea contributions.  None of these data were significantly above $Q^2 = 100$~GeV$^2$.  The Drell-Yan data were primarily in the range $16 < Q^2 < 81$~GeV$^2$ (corresponding to the dimuon mass range of $4 < M < 9$ GeV, between the $J/\psi$ and the $\Upsilon$ spectral peaks) and probed $0.01 < x < 0.2$ with the precise $x$ range shifting for each mass bin.  Only DIS data above the minimum $Q^2$ used by EPS09, 1.69~GeV$^2$, were employed in the analysis, leaving an $x$ range of $0.005 < x < 0.7$ available for fitting the modifications.  Therefore, one cannot expect great sensitivity to the individual valence and sea quark distributions.  Any small differences between the up and down valence and sea distributions in nuclear targets are due to the $Q^2$ evolution.  It is notable that only the valence quark distributions showed broad antishadowing.  In contrast,  even though they followed the same shape, the up and down sea quark distributions did not produce ratios significantly above unity and then only in the lower part of the antishadowing $x$ range.  While the sea quarks all evolve from gluons, the strange and charm sea quarks are less sensitive to the data used in the fit and thus follow the shape of the gluon ratio more directly, resulting in larger antishadowing.  Counterintuitively, the charm ratio shows more antishadowing than does the strange quark ratio.

EPPS16 \cite{EPPS16}, the successor to EPS09, was innovative in several ways. The fit used LHC data from the 2012 $p+$Pb run at a center of mass energy of 5.02~TeV.  The dijet \cite{37} and gauge boson \cite{46,48,49} data sets, while only a few points each, probed much higher $Q^2$ scales than previously possible.  Although the same fixed-target electron and muon DIS data were used for EPPS16, the group also used the extensive CHORUS data \cite{50} with $\nu$ and $\overline\nu$ beams on a lead target. (Since it also employed a fixed target, CHORUS covered a similar region in the $x$ and $Q^2$ plane as the charged lepton DIS data: $4 < Q^2 < 100$~GeV$^2$ and $0.05 < x < 0.7$).  Along with the CMS $W^\pm$ data \cite{46}, these data were sensitive to differences between the valence quark distributions as well as differences between the various sea quark distributions.  Although all these newly-incorporated data sets employed a lead target, a much heavier nucleus than relevant here, these data informed the lower $A$ results through a power-law $A$ scaling of the parameters, similar to that used in EPS09.

Because there was more information available to distinguish between the
quark distributions, each one was treated separately.
The nuclear dependence of the parameters was also
handled somewhat differently in EPPS16, adjusting the $A$ dependence 
to ensure that nuclear
effects are larger for heavier $A$.
Due to the greater number of available constraints, the number of parameters increased.  EPPS16 has 20 parameters, giving
41 total sets with one central set and 40 error sets.  The error sets and the
total uncertainty are produced the same way as in EPS09.
The sets cover the range $1.3 < Q < 10000$~GeV and
$10^{-7} < x < 1$, reaching both lower $x$ and higher $Q$ than
EPS09.

\begin{figure}
  \begin{center}
    \begin{tabular}{cc}
    \includegraphics[width = 0.45\textwidth]{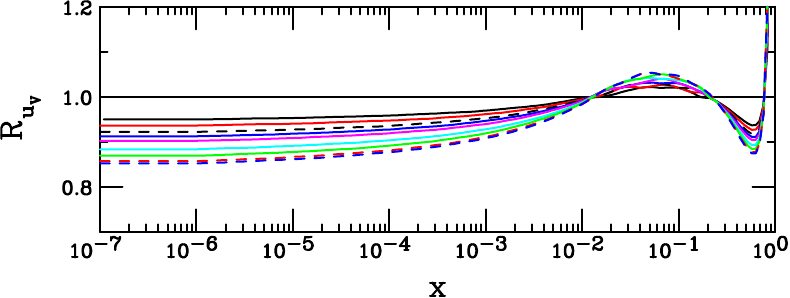} &
    \includegraphics[width = 0.45\textwidth]{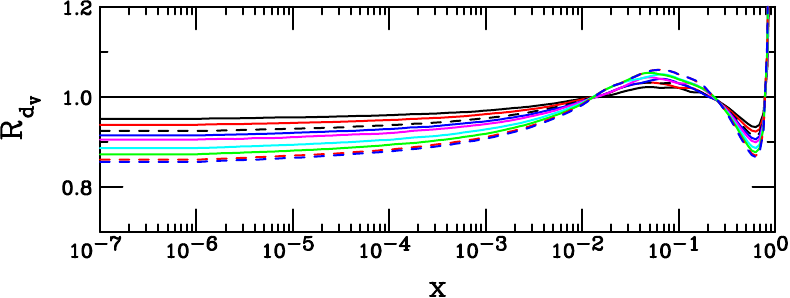} \\
    \includegraphics[width = 0.45\textwidth]{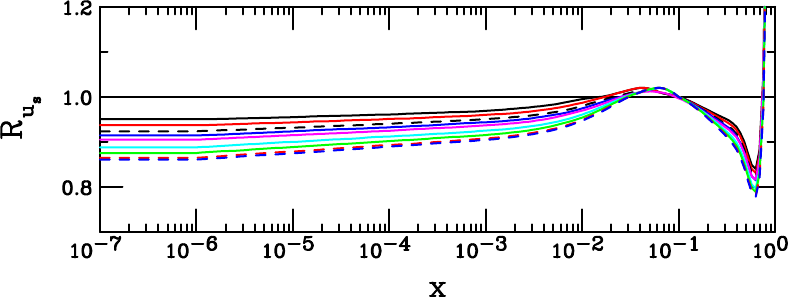} &
    \includegraphics[width = 0.45\textwidth]{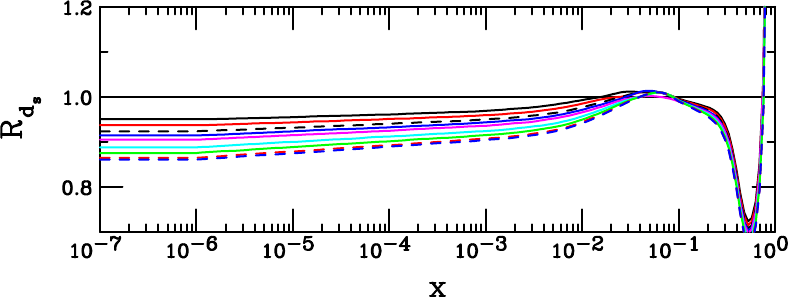} \\
    \includegraphics[width = 0.45\textwidth]{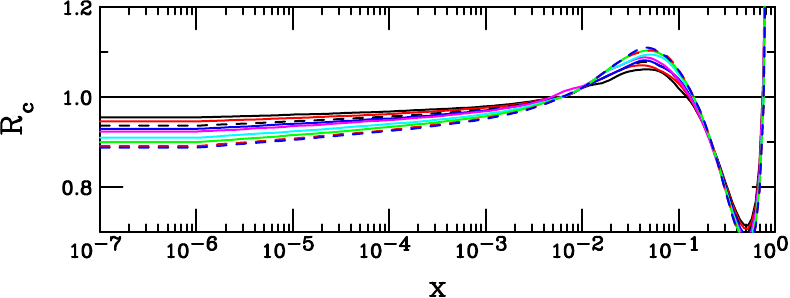} &
    \includegraphics[width = 0.45\textwidth]{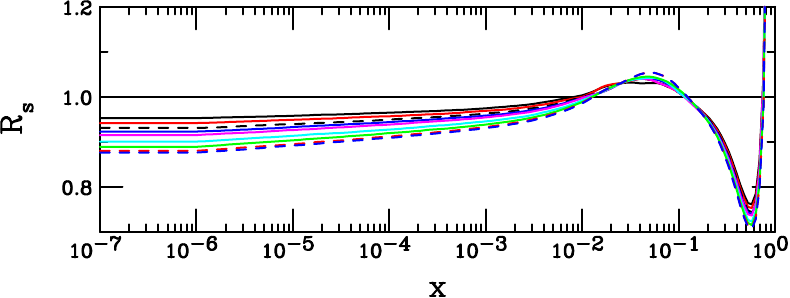} 
  \end{tabular}
\end{center}
\caption[]{(Color online)  The nuclear modification factors for EPS09 with
$u$-like quark ($u_V$, $\overline u$, and $c$) ratios on the left-hand side and
$d$-like quark ($d_V$, $\overline d$, and $s$) ratios on the right-hand side
with valence quarks at the top, the light quark sea in the center, and more
massive quarks on the bottom.  Ratios are shown for $A = 4$ (black), 6 (red),
9 (black dashed), 12 (blue), 16 (magenta), 28 (cyan),
40 (green), 56 (red dashed) and 64 (blue dashed).
Isoscalar targets are represented by solid lines;
non-isoscalar by dashed curves.
}
\label{eps09_plot}
\end{figure}

\begin{figure}
\begin{center}
    \begin{tabular}{cc}
    \includegraphics[width = 0.45\textwidth]{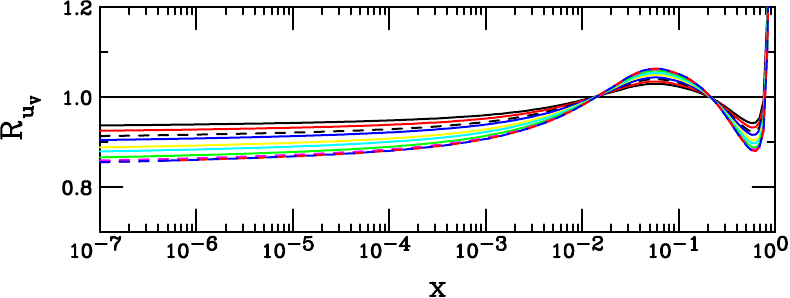} &
    \includegraphics[width = 0.45\textwidth]{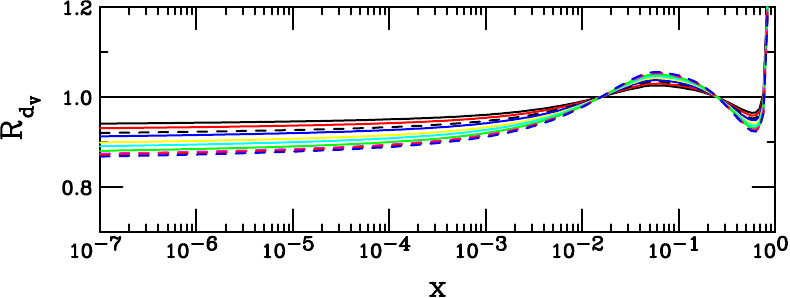} \\
    \includegraphics[width = 0.45\textwidth]{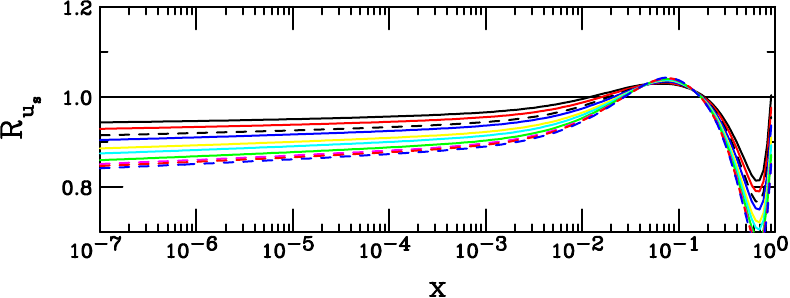} &
    \includegraphics[width = 0.45\textwidth]{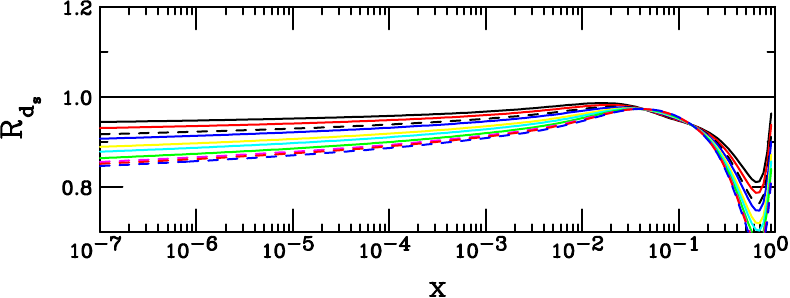} \\
    \includegraphics[width = 0.45\textwidth]{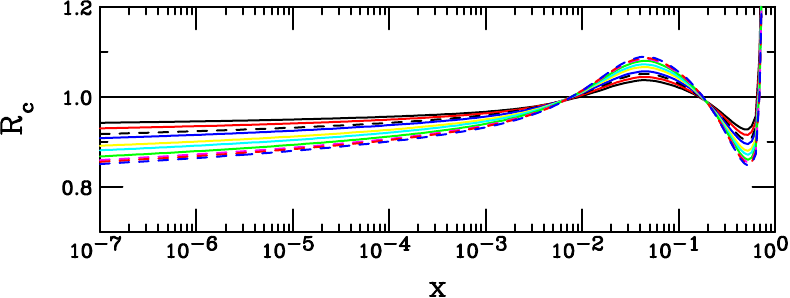} &
    \includegraphics[width = 0.45\textwidth]{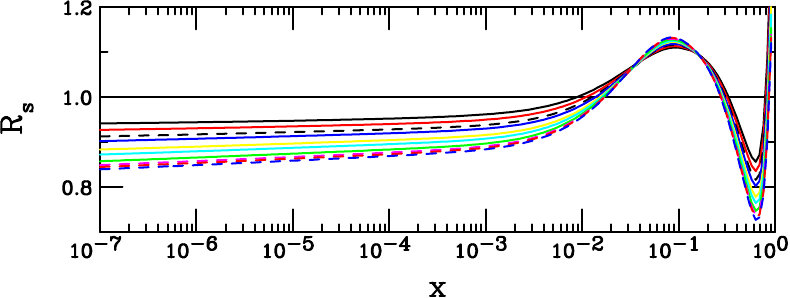} 
  \end{tabular}
\end{center}
\caption[]{(Color online)  The nuclear modification factors for EPPS16 with
$u$-like quark ($u_V$, $\overline u$, and $c$) ratios on the left-hand side and
$d$-like quark ($d_V$, $\overline d$, and $s$) ratios on the right-hand side
with valence quarks at the top, the light quark sea in the center, and more
massive quarks on the bottom.  Ratios are shown for $A = 4$ (black), 6 (red),
9 (black dashed), 12 (blue), 20 (yellow), 28 (cyan),
40 (green), 50 (magenta dashed), 56 (red dashed) and 64 (blue dashed).
Isoscalar targets are represented by solid lines;
non-isoscalar by dashed curves.
}
\label{epps16_plot}
\end{figure}

The nuclear modification ratios are shown in Figs.~\ref{eps09_plot} and
\ref{epps16_plot} for the targets included in the fits for targets up to $A = 64$, copper.  In both figures, the modifications are shown for
$Q = M_Z$.  Even for these high scales, the modification  does not vanish, although it is
 weaker than at lower scales.
Only the central sets are shown for each $A$ to highlight the general trend.  The error sets allow
for a considerably broader potential modification with the largest uncertainties at low $x$
and larger $A$.

The EPS09 sets included both LO and NLO sets with the LO set based on CTEQ61L and the NLO sets based on CTEQ6M.  The differences in the low $x$ behavior of the proton PDF sets at different orders are reflected in the shadowing ratios, especially for the gluon ratios where the LO EPS09 sets show stronger shadowing than at NLO.  EPPS16 has no LO sets.

The EPS09 valence and light sea ratios in Fig.~\ref{eps09_plot} are very similar
since they assume that, at the starting scale,
$R_{u_V}(x,Q_0^2,A) = R_{d_V}(x,Q^2,A)$ and
$R_{\overline u}(x,Q_0^2,A) = R_{\overline d}(x,Q_0^2,A)$.  Some differences
arise with $Q^2$ evolution according to the DGLAP equations giving, for example,
a more pronounced EMC effect at large $x$ for $\overline d$ ratios than
the $\overline u$.  While the strange quark ratio $R_s(x,Q_0^2,A)$ might
have started out equal to that of the light quark sea, it evolves to a larger
antishadowing than the $\overline u$ and $\overline d$ ratios.  The charm quark
sea ratios, more closely connected to the gluon ratio, show similar
significant antishadowing, as does $R_g(x,Q^2,A)$.  This may seem somewhat
counterintuitive, however, because one might expect the heavier quarks,
which enter the $Q^2$ evolution only above the quark mass scale, to be
modified less in the presence of a medium.  The $x$ range is
extended to $10^{-7}$ to show that, for the very low $x$ range, $x < 10^{-6}$,
the ratios are fixed to their value at the lowest $x$ considered by EPS09,
$10^{-6}$.

The effects of  including the CHORUS neutrino data as well as the
LHC gauge boson data in the EPPS16 sets are clearly
illustrated by comparing Fig.~\ref{epps16_plot} to Fig.~\ref{eps09_plot}.
%The additional data made it possible to distinguish between the individual quark modifications.
There are clear differences between the
valence ratios and between the sea quark ratios.  The nuclear modification is
weaker for $d_V$ than  for $u_V$ while there is no antishadowing
(quantified as a ratio larger than unity) at all for $\overline d$ in this $x$
range.  On the
other hand, the $\overline u$ ratio shows antishadowing on a level similar to
that of the valence quarks, albeit over a narrower range in $x$.  Because the
level of antishadowing in the strange quark ratio was allowed to float, it shows
stronger antishadowing than the charm ratio, closer to what might be 
%more
intuitively expected. 

As $E_\nu$ increases, the cross sections probe successively lower values of $x$,
with $E_\nu > 10^5$~GeV corresponding typically to
$x < 0.01$, marking the transition from the antishadowing to the shadowing
region where a suppression relative to deuterium should be observed.  The
modifications of $\nu$ and $\overline\nu$ for NC interactions  should be
similar because $F_{2A}$ and $F_{3A}$ are independent of the type of neutrino
initiating the interaction.  The only difference between the cross sections for
$\nu$ and $\overline \nu$ are whether $F_{2A}$ and $F_{3A}$ are summed, as for
neutrinos, or subtracted, as for antineutrinos.  In the case of CC interactions,
the differences between $\nu$ and $\overline\nu$ should be larger, especially
for non-isoscalar targets, because of the difference in valence quark content between protons and neutrons.

\begin{figure}
\begin{center}
\includegraphics[width = 0.9\textwidth]{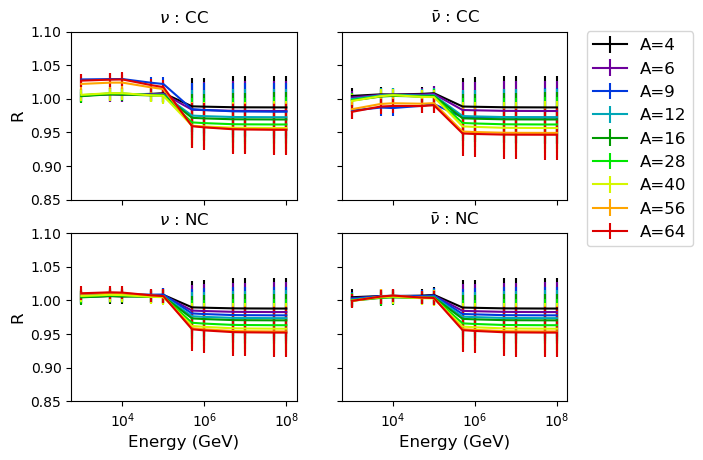}
\end{center}
\caption[]{(Color online) The nuclear modification factors for charged
current (top) and neutral current (bottom) interactions initiated by neutrinos (left) and antineutrinos (right) calculated with EPS09.
Results are shown for $A = 4$, 6, 9, 12, 16, 28, 40, 56 and 64 relative to deuterium.
}
\label{Sally_drats9}
\end{figure}

\begin{figure}
\begin{center}
\includegraphics[width = 0.9\textwidth]{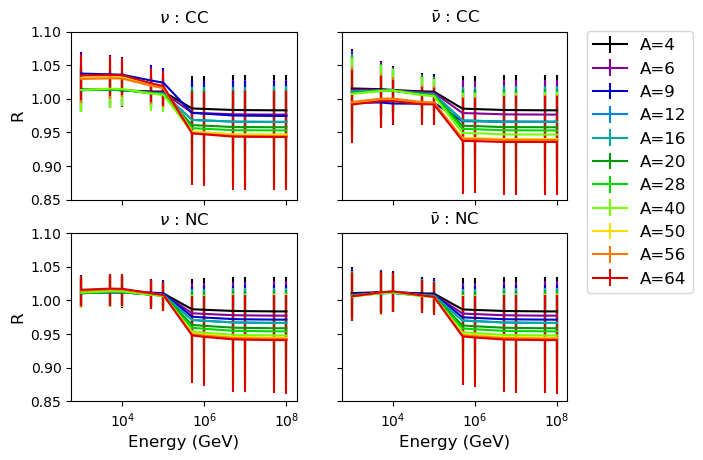}
\end{center}
\caption[]{(Color online)  The nuclear modification factors for charged
current (top) and neutral current (bottom) interactions initiated by
neutrinos (left) and antineutrinos (right) calculated with EPPS16.
Ratios are shown for $A = 4$, 6, 9, 12, 20, 28,
40, 50, 56 and 64 relative to deuterium, an isoscalar target.
}
\label{Sally_drats16}
\end{figure}

Figures~\ref{Sally_drats9} and \ref{Sally_drats16} show the ratios
$R=\sigma^{{\rm CC},{\rm NC}}(\nu (\overline \nu)A)/\sigma^{{\rm CC},{\rm NC}}(\nu (\overline \nu)d)$ as a function of neutrino energy $E_\nu$.  The curves for most of
the elements are based on global analyses including data on those elements
(including $^6$Li), but some of the heavier elements (particularly $A=50$) are
convenient intermediate nuclei for extrapolation.
The ratios are calculated for the central EPS09 and EPPS16 sets respectively.
The uncertainties are indicated
by the vertical bars for a number of energies.  The uncertainties increase with
nuclear mass number and with $E_\nu$.  Despite the additional data sets included
in the analyses, the uncertainties on EPPS16 are generally larger than on EPS09, because of the five additional fit parameters employed.  The uncertainties are also generally smaller for neutral current
interactions, probably because the $F_{3A}^{{\rm NC}}$ structure
function only has contributions from valence quarks which are better constrained
in the global analyses of the nPDFs.

All four cases show a decrease in $R$ with increasing $E_\nu$.  The fairly sharp drop between 100 and 500 TeV is due to the transition from the antishadowing region to the shadowing region.  At small $E_\nu$, there is considerable variation in shadowing for $E_\nu < 500$ TeV for CC interactions, while for NC interactions, the nuclear effects are much smaller, as we now discuss.

The charged current results display a clear separation between isoscalar and non-isoscalar targets.  The ratios for the isoscalar targets (He, Li, C, O, Ne, Si, and Ca) in charged current interactions are all somewhat higher than unity for $E_\nu \leq 10^5$~GeV, for both $\nu$ and $\overline\nu$.    However, non-isoscalar targets (Be, V, Fe and Cu) are different, with a larger enhancement for  $\nu$ and a slight suppression for $\overline\nu$ because
$N > Z$. Non-isoscalar counterparts of common nuclei (such as $^{18}$O vs.$^{16}$O or $^7$Li vs.$^6$Li) found in the crust only make up a small percentage of the total composition and thus do not contribute to our calculation. 
From Eqs.~(\ref{F2As_nuCC})-(\ref{F3As_nubCC}), we  see that the
valence distributions are weighted differently in CC interactions.  In neutrino-initiated CC interactions, $Z$ multiplies $d_V$ while $N$ multiplies $u_V$.  This is opposite for antineutrino-induced CC interactions.  When $Z = N$, there is no difference but one arises when $N > Z$.  In the case of NC interactions, Eqs.~(\ref{F2A_nc}) and (\ref{F3A_nc}), even though the couplings are different, both valence distributions are multiplied by both $Z$ and $N$, effectively diluting the effect of a non-isoscalar target, as seen in Figs.~\ref{Sally_drats9} and \ref{Sally_drats16}.
%Marker

At low $E_\nu$ the effects of isoscalar targets dominate over the effects of shadowing.   At higher $E_\nu$, the enhancement is washed out by larger contributions from sea quarks.  When $E _\nu\leq 10^5$~GeV, there should be a stronger dependence on isospin for more neutron-rich nuclei such as lead.  These results also demonstrate a non-negligible effect from the inclusion of nPDFs, beyond the difference between isoscalar and non-isoscalar targets.

\section{Nuclear effects on cross sections}

%SRK moved note about spherical symmetry to make second sentence clearer.  %RV fixed punctuation 
Neutrino telescopes measure the neutrino interaction cross section by observing neutrino absorption in the Earth \cite{Klein:2013xoa} as a function of neutrino energy $E_\nu$ and zenith angle $\theta$; spherical symmetry is assumed.  The absorption measurements can be unfolded to obtain the distribution of mass within the Earth's crust, mantle and core in different proportions, depending on $\theta$.

\begin{figure}
\begin{center}
\includegraphics[width = 0.5\textwidth]{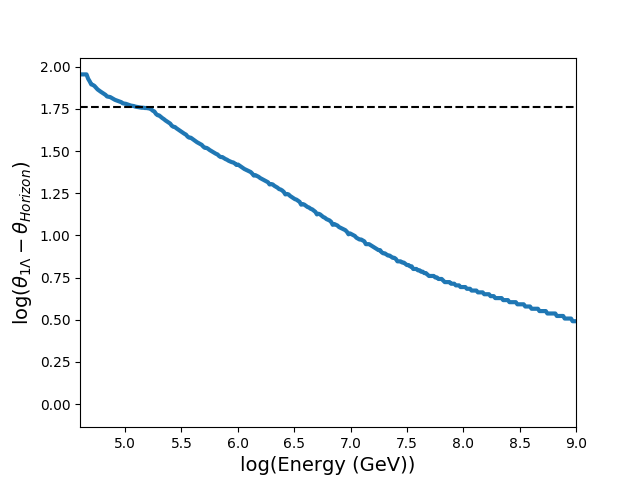}
\end{center}
\caption[]{(Color online)
  The zenith angle for which the chord length corresponds to one
  absorption length, $1 \Lambda$, as a function of energy, $E_\nu$.
  The logarith, of the angular distance between this zenith angle and the
  horizon ($\theta = 90^\circ$).  The dashed line shows the angle corresponding
  to trajectories that just touch the core-mantle boundary where the change
  in density affects neutrino absorption.
}
\label{fig:abszenith}
\end{figure}

%SRK multiple changes to paragraph, aiming for increased clarity. 
%SRK the previous version suffered from hanging clauses, etc. %RV made a few more
The neutrino interaction cross section rises with energy $E_\nu$, so the
path length, $L$, corresponding to one absorption length, $1 \Lambda$,
decreases with increasing $E_\nu$.  Path lengths of order $1 \Lambda$ are most
important for cross section measurements. For much
shorter path lengths, absorption is too small to have a significant effect.
No neutrinos survive the transit for much longer path lengths where
$L \gg 1 \Lambda$.   
When $E_\mu = 40$~TeV, a vertically-incident upward-going neutrino
(passing through the center of the Earth) travels a path length
of $\sim 1 \Lambda$.
This corresponds to a zenith angle $\theta= 180^\circ$, while horizontal
incidence is $\theta = 90^\circ$  and $\theta = 0^\circ$ corresponds to
vertically-downward incidence.   Figure~\ref{fig:abszenith} shows how the zenith
angle corresponding to a chord length equal to $1 \Lambda$
decreases with increasing neutrino energy.  The logarithm of
the angular distance from the horizon ($\theta_{\rm horizon} = 90^\circ$),
$\log(\theta_{1\Lambda} - \theta_{\rm horizon})$, is shown. Large
zenith angles are required to measure absorption at TeV-scale energies.
As the energy rises, more horizontal angles become more important.
The Earth is almost opaque to neutrinos of energies $\sim 10^9$~GeV
and experiments are most sensitive to absorption near the horizon, primarily in
the mantle and crust.

Our calculations consider $\nu_\mu$ and $\overline\nu_\mu$.  Other neutrino flavors ($\nu_e$ and $\nu_\tau$ and their antiparticles) should behave similarly, with two possible exceptions. In low-energy $\nu_\tau$ interactions the $\tau$ mass can slightly affect the kinematics. In $\overline \nu_e$ interactions with $E_\nu \sim 6.3 \times 10^6$~GeV $\overline\nu_e$ interactions with atomic electrons resonantly produce a $W^-$, the Glashow resonance \cite{Glashow:1960zz}.  At these energies, the Glashow resonance dominates the $\overline\nu_e$ cross section, complicating DIS measurements. 

To quantify the nuclear effects on neutrino absorption along different paths,  we first determine the cross sections without nuclear effects.  We then calculate the same probability with nuclear effects, using EPPS16 shadowing.  We quantify the effect by $R$, the ratio of the probabilities with and without shadowing.

We use a simplified model of the Earth, based on Table 2.2 of Ref. \cite{Earth}, which gives the elemental composition of the Earth in four regions: crust, mantle, outer core and inner core.  We use a single core region, based on an average of the inner and outer core components.  Table \ref{tab:comp} gives the percentages of oxygen, iron, magnesium, silicon, nickel, sulfur, calcium, aluminum, and sodium in each region. Reference~\cite{Earth} listed a small percentage of the composition as `other;' this percentage was not included in this study.    Some of these nuclei were not available in the EPPS16 parameterization.  To fill in these gaps, we interpolated between the $A$ values that were included in EPPS16, assuming the shadowing was proportional to $A^{1/3}$.   While most of the elements considered have the same number of protons and neutrons, the two heaviest elements, iron and nickel, have a significant neutron excess.  In addition $^{23}$Na and $^{27}$Al also have a neutron excess and are thus not isoscalar targets. 

\begin{table}[h!]
	\begin{center}		
		\begin{tabular}{|c|c|c|c|c|} % <-- Alignments: 1st column left, 2nd middle and 3rd right, with vertical lines in between
			\hline
			Element& Crust & Mantle & Core & $N-Z$  \\
			\hline			
			$^{16}$O & 46.8 &44.23& 5.40 & 0 \\
			$^{23}$Na & 2.9 & 0.27 &- & 2\\
			$^{24}$Mg & 1.3 &22.8&- & 0 \\
			$^{27}$Al & 8.0& 2.53&- & 1 \\
			$^{28}$Si & 30.8 &21.0&- & 0 \\
			$^{32}$S & 3.0&-&8.43 & 0 \\
			$^{40}$Ca & -&2.53&- & 0 \\
			$^{56}$Fe & 3.5& 6.26& 81.79 & 4 \\
			$^{59}$Ni & -& 0.2&6.70 & 3 \\
			\hline			
		\end{tabular}
		\label{tab:comp}
		%SRK added 'in percent.
	\caption{The elemental abundances in each region of the Earth's interior \cite{Earth}, in percent. The core region is an average over the inner and outer cores.  The last column, $N-Z$, shows the degree of target non-isoscalarity.}
	\end{center}
\end{table}

An average $R$, $R_{\rm region}$, is calculated in each region based on the percentage abundance of each element and using an interpolation for elements not included in EPPS16. 

\begin{figure}
	\begin{center}
		\includegraphics[width = 0.5\textwidth]{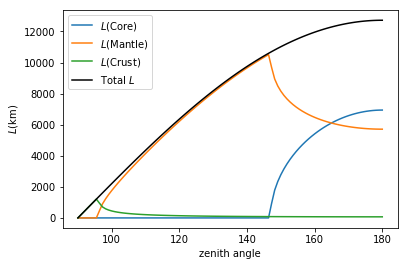}
	\end{center}
	\caption[]{(Color online)  The path length, $L$, in the crust, mantle and core based on zenith angle. At angles above $147^{\circ}$ the path length transverses the Earth's core, resulting in an observable effect on neutrino interactions, as highlighted in Fig.~\ref{fig:abszenith}.
	}
	\label{fig:pathlength}
\end{figure}

A neutrino traversing the Earth  at zenith angle $\theta$ has a path length $L$ through the three layers (crust, mantle and core), as shown in Fig.~\ref{fig:pathlength}.  Nuclear shadowing effects for each zenith angle are based on the average over the whole path length, where $R_{\rm region}$ in each region is weighted by its percentage of the total path length. 

% SRK multiple changes to correct.  "Two fluxes are obtained" isn't quite right. 
%SRK I am still a bit concerned that this paragraph may not sufficiently differentiate between what IceCube does for the analysis, and what we do in this paper.
%RV I have no real idea of the difference...
We consider CC and NC interactions separately.
CC interactions are straightforward: the neutrino interacts and disappears.  NC interactions are more complicated because the neutrino loses a fraction of its energy without disappearing.  The average fractional energy loss is 20\% at high neutrino energies.  The effect of this energy loss can be handled by treating absorption as a two dimensional matrix, relating the energy distribution of neutrinos entering the Earth to the energy distribution of detected neutrinos.  This is expressed as a relationship between two fluxes, $\Phi_{\rm in}$ a function of $E_\nu$ when the neutrino enters the Earth, and $\Phi_{\rm out}$,  a function of the detected neutrino energy \cite{Aartsen:2017kpd}.    The result can be expressed in terms of an apparent transparency, $T$, at a single energy $E_\nu$, $T=\Phi_{\rm out}(E_\nu,\theta)/\Phi_{\rm in}(E_\nu,\theta)$, but, because of energy loss by more energetic neutrinos, this apparent absorption at a given $E_\nu$ has some dependence on the assumed neutrino spectrum.    We avoid this spectral dependence here by considering only monoenergetic neutrinos.

\begin{figure}
\begin{center}
\includegraphics[width = 0.45\textwidth]{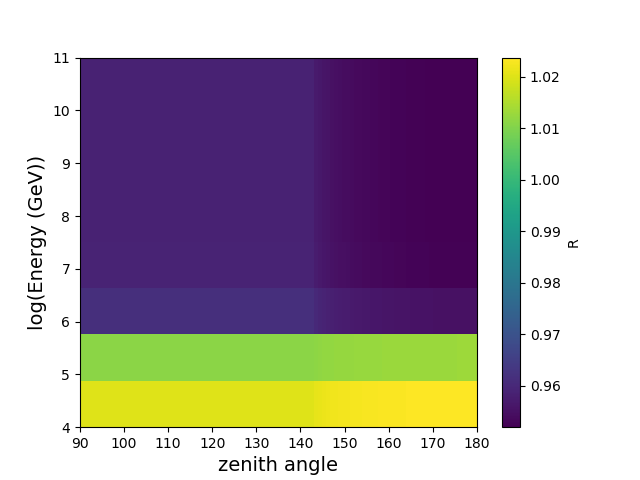}
\includegraphics[width = 0.45\textwidth]{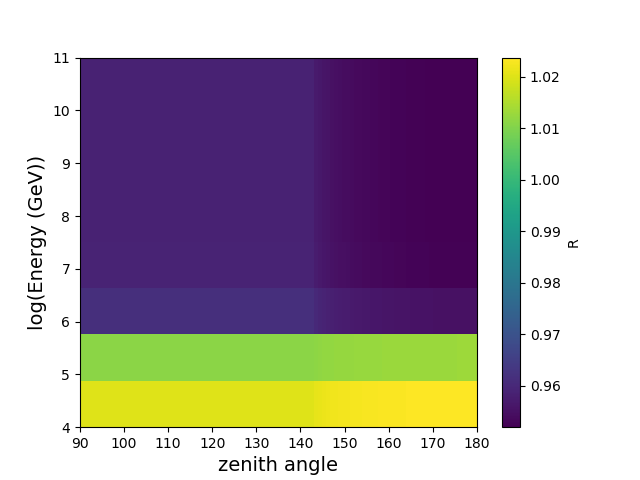}
\end{center}
\caption[]{(Color online)  The ratio of nuclear effects, $R$, for neutrinos (left) and antineutrinos (right) as a function of energy and zenith angle.   The shift from antishadowing ($R>1$) to shadowing ($R<1$) is clearly visible for energies above 500~TeV. The differences $|R-1|$ are larger for $\theta > 145^\circ$, where the Earth's core is traversed.
}
\label{fig:absorb2d}
\end{figure}

\begin{figure}
\begin{center}
	\includegraphics[width = 0.45\textwidth]{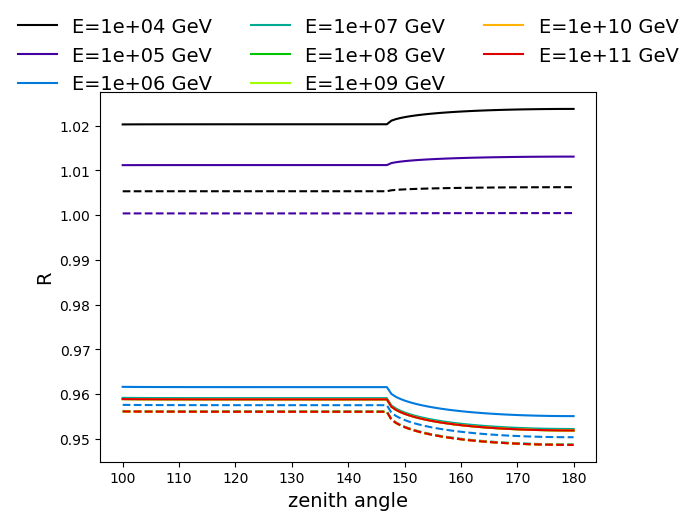}
\end{center}
\caption[]{(Color online)  The ratio of nuclear effects, $R$ for neutrinos (solid) and antineutrinos (dashed) as a function of energy and zenith angle.   The shift from antishadowing ($R>1$) to shadowing ($R<1$) is  clearly visible for energies above 500~TeV. The differences $|R-1|$ are larger for $\theta > 145^\circ$, where the Earth's core is traversed.
}
\label{fig:absorb1d}
\end{figure}

Figure~\ref{fig:absorb2d} shows $R$ as a function of neutrino energy and zenith angle, for $\nu_\mu$ and $\overline\nu_\mu$.  The shift from shadowing to antishadowing is visible at $E_\nu \sim 500$~TeV.  The effect of the Earth's core, with its high iron and likely nickel content, enhances nuclear effects for large zenith angles.  These results are shown more quantitatively in Fig.~\ref{fig:absorb1d} where $R$ is presented as a function of $\theta$ for several discrete neutrino energies.  
%At lower $E_\nu$ - this was in the Dec. 2 version, but makes no sense here.
%RV made a couple of minor tweaks
$R$ drops from about 1.02 to 0.96 for increasing neutrino energies.  With trajectories traversing the Earth's core, the spread is slightly larger due to the greater nuclear effects.  The differences between $\nu$ and $\overline\nu$ are shown to be relatively small.

The probability for a neutrino to survive passage through the Earth is 
\begin{equation}
P(E_\nu)= 1-\exp(-L(\theta)/\Lambda(E_\nu))
\end{equation}
where $L$ is the path length through the Earth for a given zenith angle $\theta$ while $\Lambda(E_\nu)\propto 1/\sigma(E_\nu)$ is the absorption length.  Nuclear effects modify the cross section so that $\sigma(E_\nu) \sim R \sigma_0(E_\nu)$ where $\sigma_0(E_\nu)$ is the cross section for an isoscalar target without shadowing.
At an energy and zenith angle where $L\approx \Lambda(E_\nu)$, the change survival probability due to the inclusion of nuclear effects, $\Delta P(E_\nu)$ is roughly comparable to $-(R-1)$.   As $L$ rises, $dP(E_\nu)/dR \approx - L/\Lambda$ so that for long chords through the Earth with a high absorption probability, $L/\Lambda\gg 1$, small changes in $R$ lead to larger changes in the absorption probability.  

The larger modification of $P(E_\nu)$ with $R$ due to shadowing could affect the interpretation of anomalous events, such as the possible $\nu_\tau$ events observed by ANITA \cite{Gorham:2018ydl}.  For example, ANITA-III event 15717147 traverses a 7,000 km chord through the Earth, corresponding to $\sim 15 \Lambda$ without nuclear effects, requiring contributions from BSM physics.    Reducing the cross section by 4\% due to shadowing roughly doubles the survival probability.  While this reduction is too small to alter the overall conclusion, a larger reduction in cross section, such as from a colored glass condensate \cite{Henley:2005ms} might allow this event to be interpreted without requiring BSM physics.  

It is important to note that, although we have focused on the best-fit values of EPS09 and EPPS16, the uncertainties are significant.   The uncertainties on these $R$ values are considerable, more than $\pm 5$~\% and thus larger than $|R-1|$.  This uncertainty will limit many quantitative analyses. 

\section{Nuclear effects on inelasticity}
\label{Sec:inelasticity}

The inelasticity of high-energy $\nu_\mu$ and $\overline\nu_\mu$ charged-current interactions on ice targets has been measured by the IceCube Collaboration \cite{Aartsen:2018vez}.  They separately measured the energy of the hadronic cascade from the struck nucleus, $E_{\rm cas}$, and the energy of the outgoing muon,
$E_\mu$, and calculated the inelasticity as $y=E_{\rm cas}/(E_\mu + E_{\rm cas})$.  There are two major reconstruction challenges to measuring neutrino inelasticity in a neutrino telescope.  First, a muon with energy above 1 TeV travels many kilometers before losing its energy, making the energy determination difficult.  Second, there are  corrections to account for missing (carried off by the neutrino) and mismeasured (hadronic instead of electromagnetic) energy.

Measurements using starting tracks, defined as a track originating in the active volume of the detector, have many applications.  In addition to inelasticity, starting tracks are important for flavor ratio measurements; determining $\overline\nu/\nu$ from the difference in inelasticity distributions; and searching for charm production in neutrino interactions \cite{Binder:2017rlx}.  Inelasticity is also relevant for $\nu_\tau$ decays since it helps determine the division of energy between the $\nu_\tau$ interaction and the $\tau$, {\it i.e.}\ the energy division between the first and second showers in double-bang ($\nu_\tau$ events with $E_\nu > 10^6$~GeV, characterized by a hadronic shower originated when the $\nu_\tau$ interacts, a $\sim 100$~m minimum ionizing track, and a subsequent cascade when the $\tau$ decays) and double-pulse (similar to double-bang but at lower energy, resulting a $\sim 10$~m minimum ionizing track) events.  Changes in the inelasticity distribution may also affect the relationship between $E_\nu$ and muon energy in through-going muon events.

\begin{figure}
	\begin{center}
		\includegraphics[width = 0.9\textwidth]{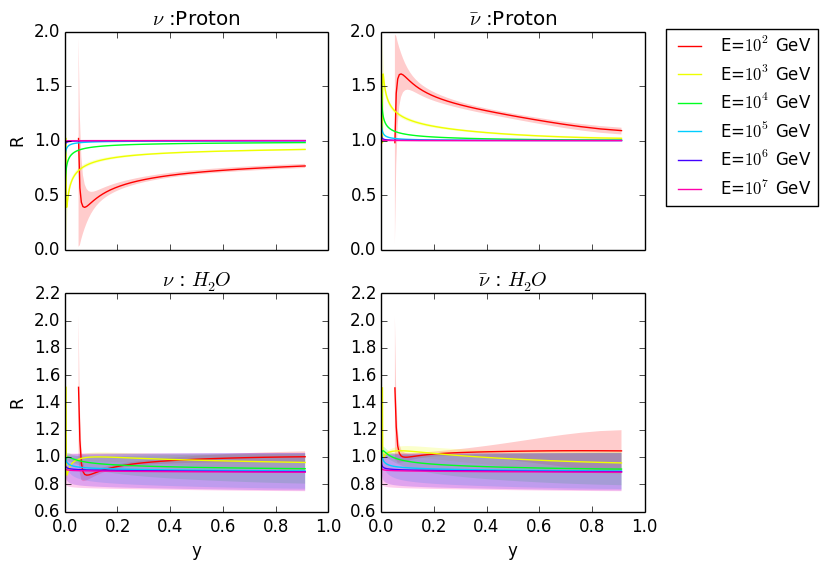}
	\end{center}
	\caption[]{(Color online) (top) The ratio $R$ as a function of inelasticity for a hydrogen target relative to deuterium. (bottom) The average ratio $R$ for H$_2$O calculated from the average of hydrogen and oxygen targets. The CT10 proton PDFs are used with EPPS16 for energies $E_\nu = 10^n$ where $n = 2-7$.
        }
	\label{fig:inelasticity10}
\end{figure}

\begin{figure}
	\begin{center}
		\includegraphics[width = 0.9\textwidth]{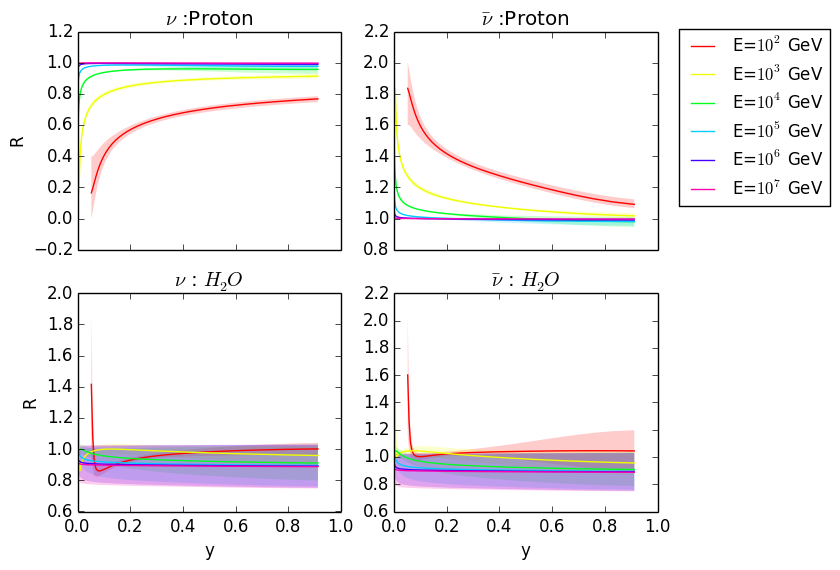}
		\end{center}
	\caption[]{(Color online) (top) The ratio $R$ as a function of inelasticity for a hydrogen target relative to deuterium. (bottom) The average ratio $R$ for H$_2$O calculated from the average of hydrogen and oxygen targets. The CT14 proton PDFs are used with EPPS16 for energies $E_\nu = 10^n$ where $n = 2-7$.
        }
	\label{fig:inelasticity14}
\end{figure}

Ice consists of H$_2$O. As we discuss, hydrogen and oxygen exhibit different nuclear effects.  While oxygen is subject to nuclear shadowing, the nuclear effects are dominated by hydrogen because it strongly violates isospin invariance.  As Figs.~\ref{fig:inelasticity10} and \ref{fig:inelasticity14} show, hydrogen has a large $|R-1|$, with a rather distinctive energy and inelasticity dependence.   The inelasticity can be related to $x$ and $Q^2$ by    
\begin{equation}
y = \frac{Q^2}{2m xE_\nu} \label{eq:ydef}
\end{equation}
where $m$ is the nucleon mass.  The inelasticity at a given $y$ and fixed $E_\nu$ is integrated over $x$ and $Q^2$.  The integral over $Q^2$ starts from a minimum $Q^2$ of 9~GeV$^2$.  The cross sections increase with $Q^2$ so that at higher $E_\nu$ the $Q^2$ integral is dominated by $Q^2\approx M_W^2$.

Figures~\ref{fig:inelasticity10} and \ref{fig:inelasticity14} show $R$ relative to an isospin invariant deuterium target for both protons and H$_2$O, using the CT10 and CT14 \cite{CT14} proton PDFs respectively with the EPPS16 nPDFs.  Results are given for values of $E_\nu$ separated by an order of magnitude for $E_\nu$ from $10^2$ to $10^7$~GeV.

In all cases, as $y$ decreases, $x$ increases.  The shape of $R$ for fixed $E_\nu$ is balance of contributions from $x$ and $Q^2$.  The $Q^2$ range at a given $y$ is constrained by the requirement that $x < 1$.  For a given $E_\nu$, as $y \rightarrow 1$, $Q^2$ can be large, near the maximum of the available range, and $x$ will remain less than unity.  On the other hand, when $y \rightarrow 0$, the denominator of Eq.~(\ref{eq:ydef}) is small so that $Q^2$ nust be near its minimum value for $x$ to be less than unity.

The choice of proton PDF is important for hydrogen at low inelasticity, where $x$ is large and the valence distributions dominate.  The top panels of the figures show $R$ for hydrogen to deuterium. With a proton target the $d$ quark distributions dominate the CC neutrino cross sections while the $u$ quark distributions dominate the antineutrino cross sections.  In the case of deuterium, the $u$ and $d$ contributions are balanced.  Thus, naively, for neutrinos $R \sim 2d/(u+d)<1$ while, for antineutrinos $R \sim 2u/(u+d)>1$.  

At large $y$, $x$ is small and decreases with increasing $E_\nu$.  Thus, in this range, the sea quark distributions dominate $R$ and $R$ approaches unity, both as $y \rightarrow 1$ and $E_\nu \rightarrow \infty$.  In models with a colored glass condensate, or strong nuclear shadowing, the cross section in this region would be reduced.

As $y \rightarrow 0$, on the other hand, the valence distributions dominate $R$ and, here, a difference between the proton PDFs appears at $y < 0.1$.  The change in the slope of $R$ in this region is due to the behavior of the $d$ valence distribution of the CT10 PDFs.  The origin of this rather abrupt change in slope in CT10 is not clear but it is absent in the CT14 sets.  We have checked the older proton PDF sets, CTEQ6M and GRV98 \cite{GRV98},  as well and the calculated $R$ values for hydrogen with these sets agree with the CT14 results.

The bottom panels of the figures show $R$ for H$_2$O.  In this case $R$ is dominated by the effect on oxygen.  As a consequence of the interplay of $x$ and $Q^2$ as a function of $y$ discussed earlier, the inelasticity curves in these figures trace the inverse of the EPPS16 shadowing ratios in Fig.~\ref{epps16_plot}.  The rise at low $y$ visible for $E_\nu = 100$~GeV is the result of the high $x$ behavior of the shadowing function; the decrease to the minimum of $R$ corresponds to the EMC region and the subsequent rise at large $y$ is the effect of antishadowing.  As $E_\nu$ increases, the EMC and antishadowing regions become more compressed at low $y$ and a decease is seen instead of a rise at larger $y$ because $x$ is in the shadowing region.  Although the effect is relatively independent of whether the interaction is initiated by $\nu$ or $\overline \nu$, there is a slightly stronger rise for $\overline \nu$-initiated interactions because of the increased antishadowing in the CC interactions.  In addition, as seen in Fig.~\ref{Sally_drats16}, the $u = u_V + \overline u$ distribution dominating antineutrino interactions includes antishadowing on both the valence and sea distributions in this region while the $d = d_V + \overline d$ dominant for neutrino interactions shows effects of antishadowing on the valence distribution but shadowing in the same $x$ region for the $\overline d$ distribution.  Finally, we note that the rather large uncertainties visible in both cases for $E_\nu = 100$~GeV arise here because, even though higher $Q^2$ values are probed at large $y$, large uncertainties remain due to the 40 error sets for EPPS16.

The uncertainties on $R$ due to the nPDF uncertainties are much larger than for the proton PDFs, on the order of 20\%.  Because the uncertainties on the nPDFs decrease with $Q^2$, as does the overall nuclear modification, the largest uncertainties are on the smallest $E_\nu$ values shown for fixed $E_\nu$.  For fixed $y$, the largest nPDF uncertainties are at small $y$ because the smallest $Q^2$ range is probed here.

Since the effect on oxygen dominates $R$ for H$_2$O, the difference between the CT10 and CT14 PDF sets does not play a significant role here.  The relative independence of the nuclear effects on the proton PDFs seen here shows that the choice of proton PDFs did not play a role in the cross section effects discussed in the previous section.   
%SRK  sentence added about possible non-linear effects.  %RV minor edits
Larger nuclear effects, due to color glass condensates or nonlinear parton dynamics, could also alter the inelasticity distribution \cite{Goncalves:2013kva}. Since $x$ and $y$ are inversely related, reductions in the cross section are likely to be most prominent at large $y$, corresponding to small $x$.

These nuclear effects have implications for neutrinos and antineutrinos with energies from 100 GeV to a few TeV.    At larger energies, the central values of $R-1$ are fairly small but with significant uncertainties due to shadowing. 

There are two key points arising from the result that $R-1 \neq 0$: the nuclear effects are large at small inelasticities and they behave differently for $\nu$ and $\overline\nu$, particularly for hydrogen.  The differences in $\nu$ and $\overline \nu$ may cancel for some effects if there is an equal admixture of $\nu$ and $\overline\nu$.   At $E_\nu \sim 1$~TeV, the flux of atmospheric neutrinos (the dominant source) is expected to be in the proportion $\nu/\overline\nu \sim 1.55$, rising to 1.75 at 100 TeV \cite{Aartsen:2017kpd}, resulting in an incomplete cancellation.  We will consider the implications of these nuclear effects on a number of different analyses.

IceCube found rather good agreement between their data and inelasticity expectations \cite{Aartsen:2017kpd}.  In their lowest energy bin, $1 < E_\nu <3$~TeV, the rise in $R$ led to an increase in low inelasticity events.  IceCube has limited low inelasticity acceptance in this energy region because of the requirement that the number of detected photons, summed over the IceCube volume, $N_{pe}$, is large enough to observe the event. The absence of a significant hadronic shower means that most of the neutrino energy is transferred to the muon and escapes the detector.  There is room in these data for nuclear effects and efforts have been made to refine the calculations of neutrino scattering rates \cite{2019PhRvD.100i1301G,2019JHEP...01..217B}.

Many other neutrino telescope measurements rely on an implicit knowledge of the inelasticity distribution because many starting-track analyses have an inelasticity-dependent inefficiency.  These analyses  also require a minimum $N_{pe}$, or other similar criteria, making them insensitive to events with $y\approx 0$.  

Inelasticity measurements have been used to measure $\nu/\overline\nu$.  At energies below about 10 TeV, $\nu$ and $\overline\nu$ have different inelasticity distributions because of the contribution from the valence quarks.  At these energies, the flux is dominated by atmospheric $\nu$ which then dominate the $\nu/\overline\nu$ measurement.   However, with a low-threshold surface air shower detector to veto neutrinos accompanied by an air shower, it might be possible to eliminate most downgoing $\nu$ from the atmosphere.   Because $\nu$ and $\overline\nu$ experience different nuclear modifications, measurements of $\nu/\overline\nu$ may require a correction for nuclear effects. 

In concert with cascades, starting tracks are also used to determine the neutrino flavor ratio.  High inelasticity starting tracks will be mistaken for cascades while low inelasticity starting tracks will be missed because of the $N_{pe}$ cut.  

Inelasticity also affects the energy that incident $\nu_\tau$ transfer to nuclear targets, affecting the energy seen in the first bang of double-bang events.

The excess of low-inelasticity events could be an unexpected background in searches for electromagnetic neutrino interactions which do not involve a nuclear recoil \cite{Seckel:1997kk,Beacom:2019pzs} and may be relevant in searches for new, BSM interaction topologies. 

\section{Conclusions}

We have examined the role of two nuclear effects: shadowing and violation of isospin invariance in ultra-high energy neutrino interactions in both the Earth and in polar ice packs.  

Antishadowing decreases the neutrino cross section by about 2\% at energies below about 500~TeV while shadowing increases it by about 4\% at higher energies.   These corrections should be included in new high-accuracy measurements of neutrino absorption in the Earth, including cross section measurements and Earth tomography.  These estimated corrections are based on standard QCD evoluion of the nPDFs; new phenomena like the colored glass condensate could perhaps lead to larger modifications. 

Nuclear effects in both hydrogen and oxygen affect the inelasticity distribution of neutrino interactions in ice, particularly at low inelasticity, in an energy range that probes quarks at large Bjorken-$x$.  The Fermi motion of nucleons in oxygen increases the number of high-momentum quarks and thus the cross section at very low $y$.  On the other hand, the difference in sign of the nuclear effects for $\nu$ and $\overline\nu$ on hydrogen will affect measurements of the ratio $\nu/\overline\nu$.  These effects are largest for $E_\nu$ from 100 GeV up to a few TeV.  At higher energies, the modifications are shifted to very low $y$ and do not contribute greatly.   

Although these central shifts are relatively small, the uncertainties are quite significant.  Until the uncertainties are reduced, they will limit the precision of many measurements.   The uncertainties rise with neutrino energy and are particularly large at the energies targeted by next-generation radio-detection systems \cite{Aguilar:2019jay}.

With the advent of higher-precision neutrino measurements \cite{Robertson:2019wfw}, it will be necessary to account for  nuclear effects in cross section and tomography measurements.  This need will only increase in larger, next-generation neutrino observatories \cite{Aartsen:2014njl,Adrian-Martinez:2016fdl,Shoibonov:2019gfj}.  
%SRK added caveat about requiring good angular resolution. $RV minor edits
The introduction of large (order 100 km$^3$) radio-pulse-based neutrino telescopes, with energy thresholds above $10^{7}$~GeV will also open the door to cross section measurements at higher energies \cite{Aguilar:2019jay,Anker:2019mnx}, provided that good angular resolution is achieved for near-horizontal events.
These experiments may also be able to measure inelasticity in $\nu_e$ interactions at very high energies by taking advantage of the LPM effect, which elongates electromagnetic showers, allowing them to be separated from the hadronic shower on the basis of Cherenkov cone widths \cite{CastroPena:2000fx} or via multiple showers \cite{Aartsen:2018vez}. 

{\bf Acknowledgements}  This work was supported in part by the National Science Foundation under grant number PHY-1307472 and the U.S. Department of Energy under contract numbers DE-AC-76SF00098 (S.R.K. and S.A.R.), DE-AC52-07NA27344 and DE-SC-0004014 (R.V).

\end{document}